\documentclass[aps,prb,twocolumn,groupedaddress]{revtex4-1}
\usepackage{graphicx}% Include figure files
\usepackage{bm}
\begin{document}

\title{Pair diffusion, hydrodynamic interactions, and available volume in dense fluids}

\author{Jeetain Mittal\footnote{Present address: Department of Chemical Engineering, Lehigh University, PA 18015}}
\email[]{jeetain@lehigh.edu}
\affiliation{Laboratory of Chemical Physics, National Institute of Diabetes and Digestive and Kidney Diseases, National Institutes
of Health, Bethesda, Maryland 20892-0520, USA}

\author{Gerhard Hummer}
\email[]{hummer@helix.nih.gov}
\affiliation{Laboratory of Chemical Physics, National Institute of Diabetes and Digestive and Kidney Diseases, National Institutes
of Health, Bethesda, Maryland 20892-0520, USA}

\date{\today}

\begin{abstract}
  We calculate the pair diffusion coefficient $D(r)$ as a function of
  the distance $r$ between two hard-sphere particles in a dense
  monodisperse suspension. The distance-dependent pair diffusion
  coefficient describes the hydrodynamic interactions between
  particles in a fluid that are central to theories of polymer and
  colloid dynamics. We determine $D(r)$ from the propagators (Green's
  functions) of particle pairs obtained from discontinuous molecular
  dynamics simulations.  At distances exceeding $\sim$3 molecular
  diameters, the calculated pair diffusion coefficients are in
  excellent agreement with predictions from exact macroscopic
  hydrodynamic theory for large Brownian particles suspended in a
  solvent bath, as well as the Oseen approximation. However, the
  asymptotic $1/r$ distance dependence of $D(r)$ associated with
  hydrodynamic effects emerges only after the pair distance dynamics
  has been followed for relatively long times, indicating
  non-negligible memory effects in the pair diffusion at short times.
  Deviations of the calculated $D(r)$ from the hydrodynamic models at
  short distances $r$ reflect the underlying many-body fluid structure,
  and are found to be correlated to differences in the local available
  volume.  The procedure used here to determine the pair diffusion
  coefficients can also be used for single-particle diffusion in
  confinement with spherical symmetry.
\end{abstract}

%Maketitle must follow title, authors, abstract, \pacs, and \keywords
\maketitle

\section{Introduction}
Pair diffusion features prominently in theories of reaction-diffusion
dynamics\cite{Zhou1991} describing processes where reactant encounters
are required, such as ligand binding and aggregation or fluorescence
quenching.  The hydrodynamic interactions quantified by the
distance-dependent diffusion coefficient are also central to the
theory and simulation of polymer dynamics, including protein folding
simulations in implicit solvent, the hydrodynamic coupling in dense
colloidal suspensions, and the function of nanomachines and bacterial
flagella.\cite{ManghiNetz:SoftMatter:2006} Considering the broad
importance of pair diffusion in theories of molecular kinetics, it may
seem surprising that little is known about the pair diffusion
coefficient and its dependence on the particle distance.  Formidable
challenges in both theory and
simulations\cite{Cukier1981,Fehder1971,Straub1990,Bocquet1997} have
resulted in often contradictory results for this fundamental quantity.

Theoretically, the pair diffusion coefficient $D(r)$ (with $r$ the
distance between two particles) has been attacked from two opposite
directions, building up from kinetic theory\cite{Cukier1981} or
projecting down from macroscopic
hydrodynamics.\cite{Happel-Brenner,ManghiNetz:SoftMatter:2006} For
$D(r)$, kinetic theory had limited success at high fluid packing
densities, largely because of the complexity of the molecular motions
in dense fluids resulting from their many-body character.  At the
other extreme, details of the molecular structure of the solvent are
ignored in estimates of the pair friction derived from macroscopic
hydrodynamics, for instance by using the Oseen or Rotne-Prager
tensors.\cite{Happel-Brenner,ManghiNetz:SoftMatter:2006} Nevertheless,
this approach has proved useful in studies of the dynamics of large
and sufficiently distant pairs of colloidal particles in a
solvent,\cite{Dufresne2000} where macroscopic hydrodynamics is
expected to apply; but it is not immediately applicable when solute
and solvent particles are of comparable size, for instance in
(aqueous) solutions of (bio)polymers.

Here, we determine the pair diffusion coefficient directly from the
simulated many-body dynamics in a dense fluid.  We focus on particles
of the same size as the solvent molecules.  This small-solute regime is of
particular relevance because, on the one hand, it allows us to
quantify hydrodynamic interactions relevant for molecular motions,
including the dynamics of (bio)polymers in solution, and, on the other
hand, it is far outside the regime where macroscopic hydrodynamics
should be expected to apply.

The paper is outlined as follows. In section I, we describe the
methodological details, including the theory to calculate the pair
diffusion tensor, the algorithm used to determine the required Green's
functions from simulation data, the simulation parameters, and the
validation procedure. We validate our method by computing the pair
diffusion coefficient for two spherical particles subject to Brownian
dynamics. In the results section
II, we first present a comparison of Green's functions obtained from
simulations against those predicted from our diffusion model, finding
excellent agreement over 8 orders of magnitude. Then we examine the
pair diffusion coefficients as a function of distance between two
particles for several fluid packing fractions, and compare the
simulation results to the predictions of macroscopic hydrodynamic
theories. Finally, we show that the position-dependent pair diffusion
coefficient is correlated to the local available volume.  In the
Appendix, we discuss the calculation of the angular pair diffusion
coefficient.

\section{Methods}
\subsection{Theory}
In the following we present the theory to calculate the
position-dependent pair diffusion tensor from simulation trajectory
data. The diffusion tensor $\mathbf{D}$ of the vector $\mathbf{r}$
between two given particles in an isotropic and homogeneous fluid has
spherical symmetry:
\begin{eqnarray}
  \label{eq:D}
  \mathbf{D}(\mathbf{r}) & = & D_\perp(r) \mathbf{e}_r\mathbf{e}_r + D_\parallel(r)\left(\mathbf{e}_\theta\mathbf{e}_\theta + \mathbf{e}_\varphi\mathbf{e}_\varphi\right)
\end{eqnarray}
where $r=|\mathbf{r}|$ is the length of the pair vector; $D_\perp(r)$
and $D_\parallel(r)$ are the scalar diffusion coefficients in the
radial and tangential directions, respectively; and $\mathbf{e}_r$,
$\mathbf{e}_\theta$, and $\mathbf{e}_\varphi$ are the orthonormal unit
vectors of the spherical polar coordinate system, with $\mathbf{e}_r$
pointing in the radial direction, and $\mathbf{e}_\theta$ and
$\mathbf{e}_\varphi$ being tangential to longitudes and latitudes,
respectively.  The Smoluchowski (or Fokker-Planck) equation describing
the diffusion of the pair vector then takes on the following form:
\begin{eqnarray}
  \label{eq:Smolu}
  \partial_t p & = & \mathrm{div} \left [\mathbf{D}(\mathbf{r})
    e^{-\beta V} \mathrm{grad} \left(e^{\beta V} p\right) \right]
\end{eqnarray}
where $p=p(r,\theta,t|r',\theta_0=0,t=0)$ is the Green's function for
a pair vector starting at a distance $r'$ and azimuthal angle
$\theta_0 = 0$, without loss of generality because of the isotropic
space (making the $\varphi$ distribution uniform); $V(r)$ is the distance-dependent free energy surface;
$\beta=(k_\mathrm{B}T)^{-1}$ is the inverse temperature; 
$\partial_t$ is the partial derivative with respect to time; and
``div'' and ``grad'' are the divergence and gradient operators in
spherical polar coordinates, respectively.

We will in the following use $x=\cos\theta$ instead of $\theta$.  Let
$P(r,x,t|r',0)$ be the Green's function in terms of this new variable.
The diffusion equation Eq.~(\ref{eq:Smolu}) then becomes:
\begin{eqnarray}
  \label{eq:Smolux}
  \lefteqn{\partial_t P = \partial_r \left[D_\perp(r)\left(\beta
        V'+\partial_r\right)P\right]}\nonumber\\ && +\frac{D_\parallel(r)}{r^2}\partial_x\left[(1-x^2)\partial_xP\right]
\end{eqnarray}
where $V'=dV(r)/dr$.  By integrating over $x=\cos\theta$ we obtain a
diffusion equation for the Green's function in the radial direction alone, with the second term
on the right hand side vanishing:
\begin{eqnarray}
  \label{eq:Dr}
 \label {smoluchowski}
  \partial_t G & = & \partial_r \left[D_\perp(r)\left(\beta V'G+\partial_rG\right)\right]~,
\end{eqnarray}
where $G(r,t|r',0)=\int_{-1}^1dx\,P(r,x,t|r',0)$ is the probability
for the pair distance to be in $(r,r+dr)$ at time $t$, starting from
$r^{\prime}$ at time 0.  As a consequence, we can treat radial
diffusion separately using standard one-dimensional diffusion,
irrespective of the angular motion.  In an appendix, we outline an
extension of the theory to the orientational diffusion of the pair
distance vector.

\subsection{Algorithm to determine pair distance diffusion coefficient}

Here we focus on the calculation of the position-dependence of the
pair-distance diffusion coefficient $D(r) \equiv D_\perp(r)$, where we
have dropped the subscript for notational simplicity.  In our
calculations of $D(r)$, we face the dual challenges that it depends on
the particle distance $r$, and that the pair dynamics becomes
diffusive only at times at which the influence is felt of the
underlying free energy surface (or potential of mean force),
$F(r)=-k_\mathrm{B}T \ln g(r) = V(r) + 2 k_\mathrm{B}T \ln r$, where
$g(r)$ is the pair correlation function of the two particles in the
fluid.  To disentangle the diffusive spread of the pair distance
distribution from the drift of the mean position as a result of the
underlying free energy surface, we use the propagator (or Green's
function) $G(r, t|r^{\prime},0)dr$.  In constructing a diffusion
model, we assume that $G$ satisfies
the Smoluchowski diffusion equation Eq.~(\ref{eq:Dr}), where  the term within
the brackets is the negative of the radial probability flux.  A spatial
discretization of the Smoluchowski equation\cite{Bicout1998} results
in a master equation that describes the pair dynamics between
neighboring intervals along $r$. The particle-pair trajectories in the
simulations are discretized by assigning pair distances into bins $i$
along $r$, and then counting the numbers $N_{ji}$ that a pair distance
is in bin $i$ at time $\tau$, and in bin $j$ at time $\tau + \Delta
t$, irrespective of its location at intervening times, with $\Delta t$
the lag time. $N_{ji}$ is symmetrized, $N_{ij}=N_{ji}$, consistent with
microscopic time reversibility.  We then find the pair diffusion
coefficient $D(r)$ that maximizes the path action of the observed
discretized path.  For the discretized diffusion model with given
$D(r_i)$ and $F(r_i)$, the path action (or likelihood) $L$ can be be
written as a product of Green's functions that are expressed in terms
of a matrix exponential.\cite{Gerhard2005}  To optimize the action
and find the diffusion model most consistent with the observed
$N_{ji}$, we infer $D(r_i)$ and $F(r_i)$ using a Bayesian
approach,\cite{Gerhard2005} with uniform priors in $\ln D(r_i)$ and
$F(r_i)$ ensuring scale invariance in time and space.

In free diffusion, one typically fits
$a+6D_0t$ (or, equivalently, $6D_0(t+\tau)$) to the mean-square
displacement, with the constant $a$ (or the time shift $\tau =
a/6D_0$) accounting for initial fast molecular motions.  Here, we
employ a similar procedure by optimizing also the time origin $\tau$
for transition counts $N_{ij}$ collected at several different lag
times $\Delta t, 2\Delta t, \ldots , k\Delta t=t$, where $t$ defines
the ``observation time.''

\begin{figure}
\centering
\scalebox{0.45}{\includegraphics*[bb= 65 65 625 275]{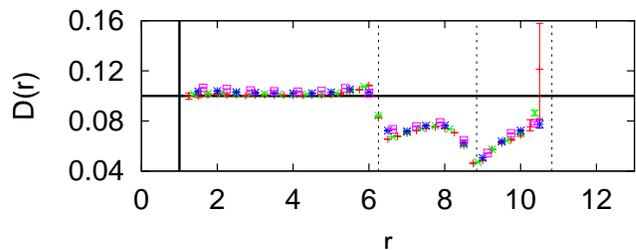}}
\caption{\label{brownian}{Pair diffusion coefficient $D(r)$ for two
    freely diffusing Brownian particles of diameter 1 with periodic
    boundary conditions.  Results for different grid sizes $\Delta r =
    0.004$ (plus), 0.032 (cross), 0.024 (star), 0.016 (square) are
    compared to the exact value $2D_0=0.1$ (horizontal line).  The
    vertical solid line marks the contact distance $r=1$. To assess
    artifacts from periodic boundary conditions, the vertical dashed
    lines mark distances $r=L/2$, $L/\sqrt{2}$, and $\sqrt{3}L/2$,
    where centered spheres touch the faces, edges, and corners of the
    cubic simulation box, respectively.}}
\end{figure}

To validate the procedure, we first run Brownian dynamics simulations
for two spherical particles of unit diameter freely diffusing with
diffusion coefficient $D_0=0.05$ in a cubic box of length $L=12.5$
under periodic boundary conditions and with reflecting boundaries at
particle contact.  By construction, in this case $D(r)=2 D_0$, which
is indeed recovered by the procedure for distances $r<L/2$
(Fig. \ref{brownian}), nearly independent of grid size $\Delta r$.
However, for $r>L/2$ and long lag times, the periodic boundary
conditions cause artifacts because in the corners of the cubic
simulation box the pair dynamics projected onto the minimum image
distance depends not only on the length of the pair vector but also on
its direction.

\section{Simulations}
To calculate $D(r)$ for a particle pair in a dense fluid, we perform
discontinuous molecular dynamics (DMD) simulations of hard sphere (HS)
fluids.  In DMD, particles follow linear trajectories between
collisions.  In a collision, the velocities of colliding particles are
changed to conserve energy and momentum.\cite{Rapaport2004}  To
simplify the notation, dimensionless quantities will be used, obtained
by appropriate combinations of a characteristic length (HS particle
diameter $\sigma$) and time scale ($\sigma \sqrt{m \beta}$, where $m$
is the particle mass). The packing fraction $\phi=\pi\rho / 6$ is
defined in terms of the particle density $\rho$.  To construct the
Green's functions, we performed DMD simulations with $N=2000$
identical HS particles.  Periodic boundary conditions were applied in
all directions.  The average self-diffusivity $D_0$ was obtained by
fitting the long-time ($t \gg 1$) behavior of the average mean-squared
displacements $\Delta {\bf r}^2$ of the particles to the Einstein
relation $\left<\Delta {\bf r}^2\right> = 6D_0t$. To minimize the
system-size dependence,\cite{Yeh2004} trajectories from simulations
with $N=10000$ particles were used to determine $D_0$, with remaining
finite-size corrections of
$\approx$1~{\%}.\cite{SigurgeirssonHeyesMolPhys2003}

\begin{figure}
\centering
\scalebox{0.4}{\includegraphics*[bb = 88 65 625 375]{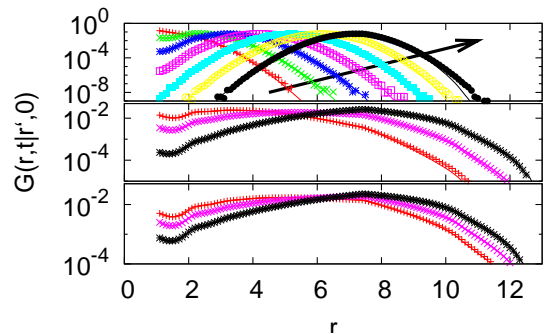}}
\caption{\label{greens}{Green's functions $G(r, t|r^{\prime},0)$ from
    simulations (symbols) and diffusion model (lines).  $G(r,
    t|r^{\prime},0)$ is shown as a function of the pair distance $r$
    at packing fraction $\phi=0.325$ for time $t=1$ (top panel), 10
    (middle panel), 20 (bottom panel). We use an observation time of $t=20$
    to obtain diffusion model parameters, combining results for lag
    times $\Delta t = 1, 2, \ldots, 20$. The arrow in the top panel
    reflects increasing $r^{\prime}=1, 2, \ldots, 7$.  }}
\end{figure}

\section{Results}

To test the applicability of the diffusion model, we compare its prediction
for the dynamics of the pair distance to actual simulation data
collected over a range of time scales.
Figure~\ref{greens} shows that diffusion quantitatively captures the
pair dynamics in the fluid.  The Green's functions $G(r,
t|r^{\prime},0)$ from the diffusion model and the results of the DMD
simulation data are found to agree over 8 orders of magnitude.  At the
shortest observation time $t=1$, we find that the Green's functions
are essentially Gaussian with position-dependent widths.  At longer
times, $t=10$ and 20, the underlying free energy surface shows its
influence, distorting the propagators away from the Gaussian form
expected for free diffusion on a flat surface.

\begin{figure}
\centering
\scalebox{0.40}{\includegraphics*[bb = 60 65 625 555]{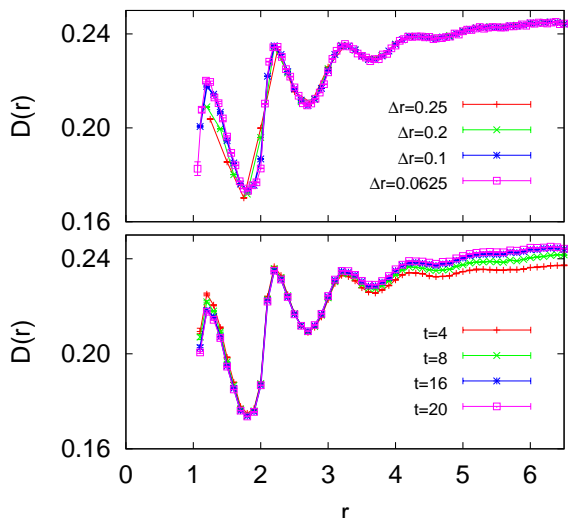}} \\
\caption{\label{param} { Dependence of $D(r)$ on diffusion model
    parameters. Pair diffusion coefficient $D(r)$ versus distance $r$
    for a hard-sphere fluid at packing fraction $\phi=0.35$ obtained
    for different (top) grid sizes $\Delta r$ (with fixed observation
    time $t=20$) and (bottom) observation times $t$ (with fixed grid size
    $\Delta r=0.1$).  The lag time is $\Delta t=1$ consistently.}}
\end{figure}

In Figure~\ref{param}, we explore the effects of the spatial grid size
$\Delta r$ and the observation time $t$ on the calculated pair
diffusion coefficient. We find that for $\Delta r \leq 0.1$, grid size
effects are negligible. Figure~\ref{param} (bottom) shows that the
effect of changing the observation time $t$ is negligible only for
shorter distances $r<3$.  In contrast, for longer distances $D(r)$ is
almost flat at a short observation time $t=4$ and does not show the
asymptotic $1/r$ dependence expected from macroscopic hydrodynamic
theory.  However, the expected $1/r$ dependence is recovered for longer
times $t$.  This result implies that the hydrodynamic coupling at
large distances is not instantaneous, such that a more accurate
diffusion model would require the inclusion of memory effects in a
frequency and position-dependent diffusion
coefficient.\cite{BryngelsonWolynesJPC1989} For $t\geq 16$ the
predictions are essentially independent of $t$. In all following
calculations, we thus use $\Delta r = 0.1$ and $t=20$.

\begin{figure}
\centering
\scalebox{0.40}{\includegraphics*[bb = 65 65 650 545]{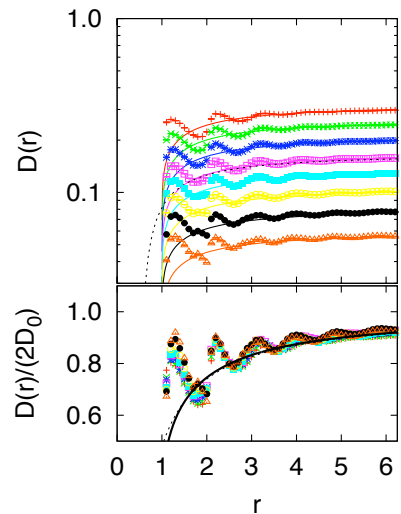}}
\caption{\label{Dr} Pair diffusion for a hard-sphere fluid. (Top)
  Calculated pair diffusion coefficient $D(r)$ versus distance $r$
  with increasing packing fraction $\phi=0.325$, 0.35, 0.375, 0.40,
  0.42, 0.44, 0.46, 0.48 (symbols, from top to bottom). Lines are the
  predictions of hydrodynamic theory (see text).  (Bottom) Normalized
  pair diffusion coefficient $D(r)/2D_0$, where $D_0$ is the
  self-diffusivity for a given $\phi$. Symbols are our calculations,
  the thick line is the exact hydrodynamic
  theory,\cite{Wolynes1976,Wachholder1972} and the dashed line is the
  Oseen approximation.}
\end{figure}

Having validated the procedure and diffusion model, we now examine the
distance-dependent pair diffusion coefficients $D(r)$ for different
packing fractions $\phi$.  Figure~\ref{Dr} (top panel) shows $D(r)$
for the HS fluid over a packing fraction range $\phi=0.325-0.48$
(symbols from top to bottom). Also shown are the predictions for
$D(r)$ from hydrodynamic theory for two spherical particles with slip
boundary conditions,\cite{Wolynes1976,Wachholder1972} as well as the
widely-used Oseen tensor correction\cite{ManghiNetz:SoftMatter:2006}
(for $\phi=0.4$; dashed line), which for the pair diffusion
coefficient is $D(r)=2D_0-k_\mathrm{B}T / (2\pi\eta r)$ where $\eta$ is
the solvent shear viscosity, taken from
Ref.~\onlinecite{SigurgeirssonHeyesMolPhys2003}.  We find that both the
exact hydrodynamic theory and the Oseen approximation (and similarly
the Rotne-Prager tensor;\cite{ManghiNetz:SoftMatter:2006} not shown)
are remarkably accurate and quantitatively reproduce the large-$r$
behavior. However, hydrodynamic predictions only qualitatively
reproduce the observed decrease in $D(r)$ near contact ($r=1$) and
lack any structure due to molecular correlations in the first- and
second-shell around a particle.

To characterize the effects of the molecular packing structure on the
pair dynamics, we plot in Fig.~\ref{Dr} (bottom panel) the normalized
pair diffusion coefficient $D(r)/2D_0$ for different packing fractions $\phi$.  As
expected from macroscopic hydrodynamics, at large distances $r$ the
$D(r)/2D_0$ data collapse onto a single curve that is well represented
by the hydrodynamic theory.  Two important observations are: (i)
$D(r)/2D_0$ is always less than 1, with pair diffusion slowed down by
``hydrodynamic interactions.''  (ii) $D(r)$ rises sharply just outside
distances of 1 and 2 particle diameters, and drops sharply just
outside $r=1.5$, and 2.5.  This strong position dependence, together
with the short-time propagator $G(r,t|r^{\prime},0)$ for Brownian
dynamics being Gaussian with mean $r=r^{\prime} + t [
D(r)\beta \partial_r F r + \partial_r D]$, implies
that at short times particle pairs just outside the first and second
shell boundary ``drift'' outward, whereas those inside the boundary
drift inward, beyond what is expected from the free energy gradient
$ \partial_r F$ alone.  The additional drift terms arise from
the large gradients in $D(r)$.  We can
understand this dynamic behavior (which does not violate microscopic
time reversibility and detailed balance!) from the many-body packing
effects.  At $r\gtrsim 2$, for instance, the interstitial space
between the two particles is likely filled by a third one, which tends
to drive the pair apart.  In contrast, at $r\lesssim 2$, the
interstitial space between the two particles is empty, and the two
particles tend to move closer together.

\begin{figure}
\centering
\scalebox{0.40}{\includegraphics*[bb = 50 65 625 560]{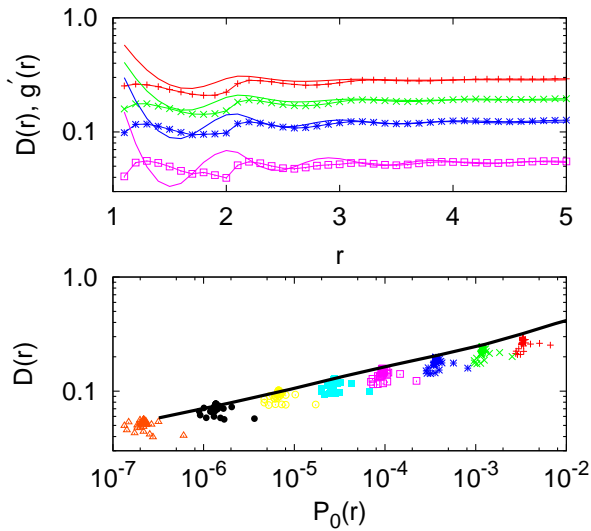}} \\
\caption{\label{gr-Dr} { Relation between fluid structure and
    dynamics. (Top) Pair diffusion coefficient $D(r)$ (symbols connected
    by lines) and scaled pair correlation function
    $g^{\prime}(r)=g(r)/a$ (lines) versus distance $r$ where $a$ is an
    arbitrary scaling factor used to match $D(r)$ and $g^{\prime}(r)$
    at large $r$ ($\phi=0.325$, 0.375, 0.42, 0.48 from top to bottom).
    (Bottom) $D(r)$ as a function of the local fractional available volume
    $P_0(r)$ (symbols; increasing packing fractions from right to
    left).  The line is $2D_0$ versus $P_0$ averaged over the entire
    system.}}
\end{figure}

To gain further insight into the observed structure in $D(r)$ and its
relation to the static structure of the fluid, we plot in
Figure~\ref{gr-Dr}a both $D(r)$ and the pair correlation function
$g(r)$.  We find that there is some correlation between the
structure in $D(r)$ and $g(r)$ except near the contact distance at
$r=1$ where these quantities are actually anti-correlated. Somewhat
counter-intuitively, this mostly positive correlation means that the
pair diffusion is actually higher in the more densely packed regions.
Similar behavior was observed for a HS fluid confined between hard
walls where the local density was found to be strongly correlated with
the local diffusion coefficient except near the
walls.\cite{Mittal2008} This behavior was found to be related to the
physics of layer formation, with the available volume, as probed by
the local test-particle insertion probability $P_0$, being largest in
the locally dense regions of space.\cite{Widom1963} A similar
argument should hold in our case of a bulk HS fluid in which purely
entropic excluded volume forces give rise to a structured $g(r)$
profile to maximize the system entropy. The local insertion
probability is given by $P_0(r) = \rho(r)/\xi= \rho g(r) /\xi$, where the activity
$\xi=\exp(\beta\mu)/\lambda^3$ is spatially invariant for an
equilibrium fluid, with $\mu$ the chemical potential, and $\lambda$
the thermal wavelength.

To test if $D(r)$ is indeed related to $P_0(r)$, we calculate $\xi$
for the different packing fractions by utilizing grand canonical
transition-matrix Monte Carlo simulations.\cite{Errington2003}
Figure~\ref{gr-Dr}b shows $D(r)$ versus $P_0(r)$ for different
$\phi$. We find that the $D(r)$ data approximately collapse onto a
curve similar to the average bulk relationship ($2D_0$ versus $P_0$)
that ignores any $r$ dependence. Therefore, at least as a rough
approximation, the local available volume can describe the pair
diffusion in this case.

\section{Concluding Remarks}
The results of this paper shed light on the microscopic origins of the
distance dependence of hydrodynamic interactions, in particular the
role of particle packing and many-body motions, and help establish a
range of validity for the assumption of macroscopic hydrodynamics in
the modeling of processes ranging from polymer dynamics to
nanomachines, colloidal dynamics, and bacterial swimming.  In
practical applications, such as the calculation of diffusional
encounter rates, the significant deviations between the calculated
pair diffusion coefficients $D(r)$ and the ideal (and widely used!)
assumption of $D(r)=2D_0 = \textit{const.}$ can result in substantial
errors, with $D(r) < 2 D_0$ consistently.  At the least one should use
a hydrodynamic theory, with both the exact theory and the Oseen tensor
giving remarkably accurate results for hydrodynamic interactions at
larger distances, and rough approximations in the regime dominated by
molecular packing near contact.

\appendix*
\section{Angular diffusion coefficient}
To treat the angular diffusion of pair distance vectors (or other
vectors in an isotropic space), we notice that the second term on the
right hand side of Eq.~(\ref{eq:Smolux}) corresponds to the angular
momentum operator in quantum mechanics.  We thus make the ansatz
$P(r,x,t|r',0)=\sum_{l=0}^\infty C_l P_l(x) q_l(r,t|r',0)$, where the
$P_l(x)$ are the Legendre polynomials of order $l$, and the
coefficients $C_l$ do not depend on $t$ and $r$. With this ansatz, we
obtain uncoupled one-dimensional evolution equations for each of the
$q_l$ (with $l=0,1, \ldots$):
\begin{equation}
  \label{eq:Dql}
  \partial_t q_l = \partial_r \left[D_\perp(r)\left(\beta
      V'q_l+\partial_r q_l\right)\right] - \frac{D_\parallel(r)}{r^2}l(l+1)q_l~.
\end{equation}
For $l=0$, this expression is identical to Eq.~(\ref{eq:Dr}); for $l>0$, these are sink (or birth-death) equations for the $q_l$, with sink terms whose strength increases quadratically with $l$, and with $D_\parallel(r)/r^2$.  That is, at long times only the distribution uniform in $x$ survives (with $P_0(x)=1$).

Expressed in terms of Dirac $\delta$-functions, the initial condition
for the Green's function is $P(r,x,t=0|r',0)=\delta(r-r')\delta(1-x)$
(where we chose the coordinate system such that the polar axis points in the direction of the pair distance vector at time zero), with normalization $\int_{-1}^1dx\int dr\, P(r,x,t|r',0)=1$.  By using the orthogonality relations of the Legendre polynomials, $\int_{-1}^1dx\,P_l(x)P_m(x)=2\delta_{lm}/(2l+1)$ with $\delta_{lm}$ the Kronecker-$\delta$, we obtain
\begin{eqnarray}
  \label{eq:P}
  P(x,r,t|r',0) & = & \sum_{l=0}^\infty \frac{2l+1}{2} P_l(x) q_l(r,t|r',0)
\end{eqnarray}
where the $q_l$ satisfy Eq.~(\ref{eq:Dql}) with initial conditions $q_l(r,0|r',0)=\delta(r-r')$.

For the sake of completeness, we also sketch an algorithm to obtain
the distance-dependent radial and angular diffusion coefficients
$D_\perp(r)$ and $D_\parallel(r)$ from simulation data (or,
equivalently, from experimental data, such as those obtained in
colloidal-particle tracking experiments).
\begin{enumerate}
\item Use counts of transitions $N_{ji}$ from bins $i$ to $j$ in the
  radial direction only (irrespective of the angular motion) as input
  in the algorithm\cite{Gerhard2005} described above to calculate the
  one-dimensional position-dependent diffusion coefficients
  $D_\perp(r)$, and the potential of mean force $V(r)$.
\item Determine counts $N_{j\alpha,i}$ for transitions from bin $i$ in
  the radial direction to bin $j,\alpha$ in a two dimensional
  histogram.  Radial bins are indexed by $j$, and angular bins by
  $\alpha$ according to the cosine of the azimuthal angle,
\begin{eqnarray}
  x(t) & = & \cos \theta(t) = \frac{\mathbf{r}(t)\cdot\mathbf{r}(0)}{|\mathbf{r}(t)||\mathbf{r}(0)|}
\end{eqnarray}
(with $\theta(0)=0$ and $x(0)=1$ by definition of the coordinate system).
\item With $D_\perp(r)$ and $V(r)$ already determined in the first
  step, the Green's function Eq.~(\ref{eq:P}) can be calculated for a
  given estimate of $D_\parallel(r)$ from a spatially discretized
  version\cite{Bicout1998} of the sink equations, Eq. (\ref{eq:Dql}).
  With this Green's function, one can again use a Bayesian inference
  procedure (or maximum-likelihood method) to estimate the
  $D_\parallel(r)$ (on lattice points halfway between the bin centers)
  that is most consistent with the observed transition counts
  $N_{j\alpha,i}$.
\end{enumerate}
Note that the infinite sum over $l$ in Eq.~(\ref{eq:P}) has to be
truncated in practical calculations.  Note further that the same
algorithm can also be used to determine the diffusion coefficients of
a single particle in confinement with spherical symmetry.

\begin{acknowledgments}
  We thank Dr. Attila Szabo for many helpful discussions. This research
  was supported by the Intramural Research Program of the NIH,
  NIDDK, and utilized the high-performance computational
  capabilities of the Biowulf PC / Linux cluster at the National
  Institutes of Health, Bethesda, MD (http://biowulf.nih.gov).
\end{acknowledgments}

%\bibliography{pairdiff}
%\bibliography{../../../../../../all}
%\bibliography{paper}

\end{document}